\documentclass[a4paper,twocolumn,%tightenlines,
english,aps,pre,floatfix,showpacs]{revtex4}
\usepackage[T1]{fontenc}
\usepackage[latin1]{inputenc}
\usepackage{amsmath}
\usepackage{babel}
\usepackage{graphics}
\usepackage{amssymb}


\begin{document}
 
\title{Finite-size investigation of scaling corrections in the \\ 
square-lattice three-state Potts antiferromagnet}

\author{S.L.A. \surname{de Queiroz}}

\email{sldq@if.ufrj.br}

\affiliation{Instituto de F\'\i sica, Universidade Federal do
Rio de Janeiro, Caixa Postal 68528, 21945-970
Rio de Janeiro RJ, Brazil}

\date{\today}

\begin{abstract}
We investigate the finite-temperature corrections to scaling in the 
three-state square-lattice Potts antiferromagnet, close to the critical
point at $T=0$. Numerical diagonalization of the transfer matrix on
semi-infinite strips of width $L$ sites, $4 \leq L \leq 14$, yields
finite-size estimates of the corresponding scaled gaps, which are 
extrapolated to $L\to\infty$.
Owing to the characteristics of the quantities under 
study, we argue that the natural variable to consider is $x \equiv 
L\,e^{-2\beta}$. 
For the extrapolated scaled gaps we show that square-root corrections, in 
the variable $x$,
are present, and provide estimates for the numerical values of
the amplitudes of the first-- and second--order correction terms, for both
the first and second scaled gaps. 
We also calculate the third scaled gap of the 
transfer matrix spectrum at $T=0$, and find an extrapolated value of the
decay-of-correlations exponent, $\eta_3=2.00(1)$. This is at odds
with  earlier predictions, to the effect that the third relevant operator 
in the problem would give $\eta_{{\bf P}_{\rm stagg}}=3$,
corresponding to the staggered polarization.

\end{abstract}
\maketitle
%\tightenlines
\section{INTRODUCTION}
\label{intro}

The three-state Potts antiferromagnet on a square lattice exhibits a
second-order phase transition at $T=0$, with distinctive properties. 
Among these is the exponential divergence of quantities such as the 
correlation length and staggered susceptibility. 

While earlier studies  agreed in pointing to a temperature dependence of 
the bulk correlation length in the form
\begin{equation}
\xi_\infty(\beta) \sim \beta^p\exp(v \beta^x)\ ,\qquad \beta \equiv 
\frac{1}{k_BT}\ ,
\label{eq:xiconj}
\end{equation}
different conjectures were advanced for the values of $p$, $v$ and $x$,
mostly on the basis of numerical work.
In particular, the value of $x$ was variously estimated as 1.3
(transfer-matrix results~\cite{Night82} analysed by the Roomany-Wyld
approximant~\cite{rw}, and Monte Carlo work~\cite{Wang90}); 
3/4 (conformal invariance arguments coupled with an analysis of the
eigenvalue spectrum of the transfer matrix~\cite{saleur91});
and 1 (further Monte Carlo work~\cite{Ferreira95}).
Later studies~\cite{gof4}, applying crossover arguments
to transfer-matrix data taken with
an external field $H$, near the critical point $T=H=0$, 
gave $x=1.08 \pm 0.13$. Additional evidence compatible with $x=1$,
and $v=2$,  was  found via extensive Monte Carlo 
simulations~\cite{feso99}.
In this latter reference it was argued that, although $p \simeq 1$
gave the best fits  to numerical data, such a logarithmic correction to 
the dominant behaviour was difficult to justify on theoretical grounds;
also, a value of $p=0$ could be made to fit the data, albeit with 
poorer quality than for $p=1$.

A substantial step towards fuller understanding of the critical properties
of the model was given in Ref.~\onlinecite{cjs01}. Through a mapping 
to the six-vertex model, where the most relevant excitations are vortices, 
the authors were able to find that
the bulk correlation length diverges as above, with the following exact 
values for the corresponding parameters: $x=1$, $v=2$, $p=0$. Further, 
they established the form of the leading corrections to scaling, so that
\begin{equation}
\xi_\infty(\beta) = A\,e^{2 \beta}\,\left(1+b\,\beta\,e^{-\beta}+ 
\cdots\right)\ .
\label{eq:xiex}
\end{equation}
The value $b=-6.65 \pm 0.11$ was calculated, upon 
consideration of the stiffness constant of a related model where
non-vortex defects are the main excitations. Similar results were
derived for the bulk staggered susceptibility.
Finally, it was shown that
the data of Ref.~\onlinecite{feso99} are compatible with the 
predictions just quoted. The 
effects previously ascribed to logarithmic corrections could be explained 
once the corrections to scaling, in the form and sign predicted, were 
taken into account. The constant in Eq.~(\ref{eq:xiex}) was 
fitted to $A=0.121(3)$~\cite{cjs01}, close to the earlier estimate 
$A \simeq 0.11 - 0.12$ for $p=0$ in Ref.~\onlinecite{feso99}.

Data in Ref.~\onlinecite{feso99} were taken for $2.0 \leq \beta 
\leq 6.0$ (corresponding to $5 \lesssim \xi_\infty(\beta)\lesssim 
20,000$), on $L \times L$ lattices with $32 \leq L \leq 1536$. Thus,
in most cases extrapolation procedures were used to estimate the 
$L \to \infty$ limiting values of the quantities of interest.

On account of the exponential divergences,
the error bars associated to extrapolated quantities turned out
to increase steeply for lower temperatures. For example (see Table 4
of Ref.~\onlinecite{feso99}), the estimate of $\xi_\infty$ starts
with a relative error of $1\%$ at $\beta=2.5$, which slowly grows to
$3\%$ at $\beta=5.2$ but then reaches $16\%$ at $\beta=5.9$, and
$33\%$ at $\beta=6.0$. Therefore, the picture
at the high-$\beta$ end of the fits to theory in Ref.~\onlinecite{cjs01}
is less than entirely clear. 

Our main purpose here is to complement the test of 
Ref.~\onlinecite{cjs01}, by 
means of transfer-matrix data generated on $L \times \infty$ strips 
of the square lattice. Being essentially exact results of numerical 
diagonalization, our data do not suffer from the fluctuations
intrinsic to Monte-Carlo studies, allowing one to reach arbitrarily
large $\beta$, in principle; instead, owing to limitations in 
the largest strip width accessible (we used $4 \leq L \leq 14$, $L$ even, 
with periodic boundary conditions across), the most important 
potential source of uncertainties is the $L \to \infty$ extrapolation. 
This drawback is somewhat mitigated by the rather smooth behaviour of 
finite-$L$ data sequences, as shown below.

\section{Strip scaling and finite-size corrections}
\label{ss+fsc}

The choice of quantities to investigate is, in part, dictated 
by
specific features of the strip geometry; here we have chosen to calculate
the first and second scaled gaps:
\begin{equation}
\eta_i= \lim_{L \to \infty}\,\frac{L}{\pi\xi_i}\ , \quad
  \xi_i^{-1}=\Lambda_0-\Lambda_i\ ,\quad i=1,2
\label{eq:scgd}
\end{equation}
where $e^{\Lambda_j}$ are the ($L$--dependent) largest eigenvalues of the 
transfer matrix.
At the critical point $\beta=\infty$, conformal invariance~\cite{cardy}
states that these quantities give the respective decay-of-correlation 
exponents; in the present case, the lowest gap $i=1$ is related to the
staggered magnetization, with associated exponent $\eta_{\rm\,stagg}=1/3$,
while $i=2$ gives the uniform magnetization decay, 
$\eta_{\rm\,u}=4/3$~\cite{dN82,saso98}. The next relevant operator is
related to the staggered polarization~\cite{dN82,saso98}, and will be 
briefly discussed in connection with scaling of the third gap ($i=3$),
in Section \ref{stg}.

For finite $\beta$ one is off criticality, thus the $\eta_i$ above are not
to be interpreted as exponents; nevertheless, they are quantities whose
difference from the {\em bona fide} $\beta=\infty$
exponents is expected to depend on powers of the (suitably 
defined) distance to the critical point.  
    
According to finite-size scaling~\cite{barber} one must have, with 
$\xi_\infty(\beta) \equiv \xi_1(\beta,L=\infty)$ given by 
Eq.~(\ref{eq:xiex}):
\begin{equation}
\frac{L}{\pi\xi_i(\beta,L)} =f_i\left(\frac{L}{\xi_\infty(\beta)}\right)
\ .
\label{eq:fss}
\end{equation}
Since the finite--{\it size} corrections here usually are of larger
magnitude than the finite--{\it temperature} ones, we shall only
take into account the dominant temperature dependence of 
$\xi_\infty(\beta)$, that is, we shall write
\begin{equation}
\frac{L}{\pi\xi_i(\beta,L)}=f_i(x)\ ,\quad x \equiv 
L\,e^{-2\beta}\ .
\label{eq:fss2}
\end{equation}
On the other hand, the incorporation of the finite--$L$ effects will be 
done phenomenologically, as explained in the following.

At $T=0$ (that is, $x \equiv 0$), very good convergence of finite-width 
estimates ($\eta_i(L)$) of 
$\eta_{\rm \, stagg,\,u}$ towards the exact results ($\eta_i$) is 
attained  by assuming corrections of the form:    
\begin{equation}
\eta_i(L)=\eta_i +\frac{a_{i\,0}}{L^2}+\frac{b_{i\,0}}{L^4} + \cdots\ .
\label{eq:corr0}
\end{equation}
These so-called `analytical' corrections, in powers of $L^{-2}$, 
are expected to occur
for any theory on a strip geometry, as they are related to the conformal 
block of the identity
operator~\cite{car86}. They will be the main corrections, provided that 
no other irrelevant operator with
a low power arises (as is the case for the three-state Potts 
{\em ferromagnet}~\cite{nien82,sldq00} where an $L^{-4/5}$ term is 
present).
In order to illustrate how Eq.~(\ref{eq:corr0}) works, and to give 
readers the opportunity to try their own extrapolation procedures,
Table~\ref{table1} gives our finite--$L$ estimates of 
$\eta_{\rm\,stagg}$ and $\eta_{\rm\,u}$, together with their 
respective extrapolations via equal-weight least-squares fits of data
(we systematically discard $L=4$ data). Error bars quoted are the
standard deviations of the estimated intercepts at $L^{-1}=0$, as given by
standard least-squares fitting procedures.  
\begin{table}
\caption{Finite--$L$ and extrapolated estimates of $\eta_{\rm stagg}$, 
$\eta_{\rm u}$. The latter  are the results of equal-weight fits of data 
for $L=6$, $8 \cdots 14$ respectively to a single-power ($L^{-2}$) 
correction (Extr.~1) and to  
Eq.~(\protect{\ref{eq:corr0}}) (Extr.~2). 
}
\vskip 0.1cm
 \halign to \hsize{\hskip1.5truecm\hfil#\hfil&\hfil#\hfil&\hfil#\hfil\cr
     L        &\ $\eta_{\rm stagg}$ &\ $\eta_{\rm u}$ \cr
     4        &\  0.308785582 &\ 1.47544318 \cr
     6        &\  0.321556256 &\ 1.39168002 \cr
     8        &\  0.326473031 &\ 1.36528410 \cr
     10       &\  0.328860921 &\ 1.35352975 \cr
     12       &\  0.330193867 &\ 1.34726477 \cr
     14       &\  0.331011103 &\ 1.34352745 \cr
     Extr. 1  &\  0.3331(1)     &\ 1.3324(3) \cr
     Extr. 2  &\ 0.333303(5) &\  1.333347(2) \cr 
     Exact    &\     1/3     &     4/3  \cr}
\label{table1}
\end{table}
Before going further, it
must be stressed that this structure of corrections to scaling is,
in principle, specific to strip  geometries~\cite{car86}; thus it
is not surprising that different results (namely, corrections to 
$\xi/L$ given by  $B\,L^{-1}+C\,L^{-5/3}$) have been found for this same
model, also at $T=0$, on fully finite $L \times L$ 
lattices~\cite{saso98}.

In order to disentangle the finite--temperature corrections (to bulk 
behaviour) which are of
interest here, we shall {\it assume} that, for fixed $x=L\,e^{-2\beta}$
one can still write
\begin{equation}
\eta_i(L,x)=\eta_i(x) +\frac{a_i(x)}{L^2}+\frac{b_i(x)}{L^4} + \cdots\ .
\label{eq:corrx}
\end{equation}
where $\lim_{x \to 0} a_i(x)=a_{i\,0}$ and similarly for the other 
$x$--dependent quantities. 
In this way we expect to account for the explicit $L$--dependence of
our finite-width results, being left only with that 
given through the argument of Eq.~(\ref{eq:fss}), which is 
intrinsic to scaled gaps. 

We illustrate the validity of the smoothness assumption just made, by 
displaying
in Table~\ref{table2} our data for the largest value of $x$ used (see 
below), $x_{\rm max}=0.04096$.

\begin{table}
\caption{Finite--$L$ and extrapolated estimates of $\eta_1$, 
$\eta_2$, for $x=0.04096$. 
The latter  are the results of equal-weight fits of data 
for $L=6$, $8 \cdots 14$ respectively to a single-power ($L^{-2}$) 
correction (Extr.~1) and to  
Eq.~(\protect{\ref{eq:corrx}}) (Extr.~2). 
}
\vskip 0.1cm
 \halign to \hsize{\hskip1.5truecm\hfil#\hfil&\hfil#\hfil&\hfil#\hfil\cr
     L        &\ $\eta_1$ &\ $\eta_2$ \cr
     4        &\  0.395983934 &\ 1.64309174 \cr
     6        &\  0.402292849 &\ 1.58471908 \cr
     8        &\  0.404971786 &\ 1.55458116 \cr
     10       &\  0.406303355 &\ 1.53956140 \cr
     12       &\  0.407011253 &\ 1.53059628 \cr
     14       &\  0.407389676 &\ 1.52458596 \cr
     Extr. 1  &\  0.4086(1)     &\ 1.512(1) \cr
     Extr. 2  &\  0.40862(6)   &\ 1.5090(6)  \cr 
}
\label{table2}
\end{table}
Comparison with Table~\ref{table1} shows that, although standard 
deviations have increased by roughly two orders of magnitude, they
still keep within quite reasonable bounds, giving credence to
the smoothness assumption underlying Eq.~(\ref{eq:corrx}) for all 
intermediate--$x$ values used here.

\section{Finite-temperature corrections}
\label{ftc}

In the analysis of the extrapolated (bulk) quantities, we shall check for 
corrections to scaling in the $x$ variable, that is,
\begin{equation}
\eta_i(x)=\eta_i(0)+C_i\,\sqrt{x} + D_i\,(\sqrt{x})^2 + \cdots\ .
\label{eq:bcorrx}
\end{equation}  
Note that a literal translation of Eq.~(\ref{eq:xiex}) would suggest
that the corrections in Eq.~(\ref{eq:bcorrx}) should depend on 
$\sqrt{x}\,\ln x$ rather than $\sqrt{x}$ alone; however, consistently with 
the argument used in establishing Eq.~(\ref{eq:fss2}), here we 
shall deal only with the dominant terms.
 
We have taken $x$--values decreasing by powers of two, from
$x_{\rm max}=0.04096$ to $x_{\rm min}=x_{\rm max}/2^{27}=3.05176 \times
10^{-10}$. Using, as a first-order approximation, $\xi_\infty(\beta)=
A\, e^{2\beta}$ with $A=0.121(3)$~\cite{cjs01}, the above values of $x$
correspond to the range $\beta=2.29$, 
$\xi_\infty \simeq 10$ ($x=x_{\rm max}$, $L=4$) to $\beta=12.27$, 
$\xi_\infty \simeq 5.4 \times 10^{6}$ ($x=x_{\rm min}$, $L=14$). 
The lower limit was set by determining when the difference between
the central estimates
$\eta_i^{\rm ext}(x_{\rm min})-\eta_i^{\rm ext}(0)$ became of the 
same order as
the standard deviation of either extrapolated quantity (see  
Table~\ref{table3} below, where one sees that, although this criterion 
has been followed strictly for $\eta_2$, the three smallest--$x$ entries
for $\eta_1$ are in fact below the threshold; however, by performing 
analyses with and without the corresponding data, we have checked that 
this is of no great import to our conclusions).  

\begin{table}
\caption{Extrapolated values of $\eta_1(x)$, $\eta_2(x)$. For each $x$ 
they are the result of an equal-weight fit of data for $L=6$, $8 \cdots 
14$ to  Eq.~(\protect{\ref{eq:corrx}}). 
}
\vskip 0.1cm
 \halign to \hsize{\hskip1.5truecm\hfil#\hfil&\hfil#\hfil&\hfil#\hfil\cr
     x      &\ $\eta^{\rm ext}_1$ &\ $\eta^{\rm ext}_2$ \cr
  0.          &\ .333303(5) &\  1.333347(2) \cr 
  3.05176E-10 &\ .333305(5) &\  1.333352(3) \cr  
  6.10352E-10 &\ .333306(5) &\  1.333354(3) \cr  
  1.2207E-09  &\ .333306(5) &\  1.333358(3) \cr
  2.44141E-09 &\ .333308(5) &\  1.333362(3) \cr
  4.88281E-09 &\ .333309(5) &\  1.333369(3) \cr
  9.76563E-09 &\ .333311(5) &\  1.333378(3) \cr
  1.95313E-08 &\ .333315(5) &\  1.333391(3) \cr
  3.90625E-08 &\ .333319(5) &\  1.333409(4) \cr
  7.8125E-08  &\ .333326(5) &\  1.333435(4) \cr
  1.5625E-07  &\ .333335(5) &\  1.333471(5) \cr
  3.125E-07   &\ .333347(5) &\  1.333523(6) \cr
  6.25E-07    &\ .333366(4) &\  1.333596(7) \cr
  1.25E-06    &\ .333392(4) &\  1.333701(10) \cr
  2.5E-06     &\ .333429(4) &\  1.333849(13) \cr
  5.E-06      &\ .333483(2) &\  1.334060(17) \cr
  1.E-05      &\ .333560(1) &\  1.334363(22) \cr
  2.E-05      &\ .333673(2) &\  1.334796(31) \cr
  4.E-05      &\ .333839(4) &\  1.33542(4) \cr
  8.E-05      &\ .334087(8) &\  1.33634(6) \cr
  .00016      &\ .334465(13) &\  1.33768(8) \cr
  .00032      &\ .335054(21) &\  1.33969(11) \cr
  .00064      &\ .335999(32) &\ 1.34276(15) \cr
  .00128      &\ .33756(5)   &\ 1.34754(22) \cr
  .00256      &\ .34025(6)   &\ 1.35523(30) \cr
  .00512      &\ .34507(9)   &\ 1.3681(4) \cr
  .01024      &\ .35409(11)  &\ 1.3904(5) \cr
  .02048      &\ .37183(12)  &\ 1.4309(6) \cr 
  .04096      &\ .40862(6)   &\ 1.5090(6) \cr}
\label{table3}
\end{table}

Table~\ref{table3} also shows
that, although our extrapolations are very {\em precise}, owing to the 
remarkably smooth variation of data against $L$, they seem
to suffer from a slight lack of {\em accuracy}. Indeed, for $x=0$
our central estimates $\eta^{\rm ext}_i$ stand  respectively 6 and 8 
standard 
deviations away from the known exact values for $\eta_1$ and $\eta_2$.
We ascribe this effect to {\em systematic} errors coming from: (i) the
shortness (in $L$) of our data series, and (ii) higher-order corrections, 
ignored  in Eq.~(\ref{eq:corrx}). Since, at least for $x=0$, such errors 
amount to small differences
in the central estimates (respectively $-0.01\%$ and $+0.001\%$ for 
$\eta_1$ and $\eta_2$) relative to exact values, and assuming this 
scenario to  carry over, continuously and smoothly, to $x \neq 0$,
we shall do as follows. In Eq.~(\ref{eq:bcorrx}), for instance,
we shall use $\eta^{\rm ext}_i(0)$ instead of the exact $\eta_i(0)$;
this way we expect systematic errors to cancel to a
large extent, when considering the difference  $\eta^{\rm 
ext}_i(x)-\eta^{\rm ext}_i(0)$.

\begin{figure}
{\centering \resizebox*{3.4in}{!}{\includegraphics*{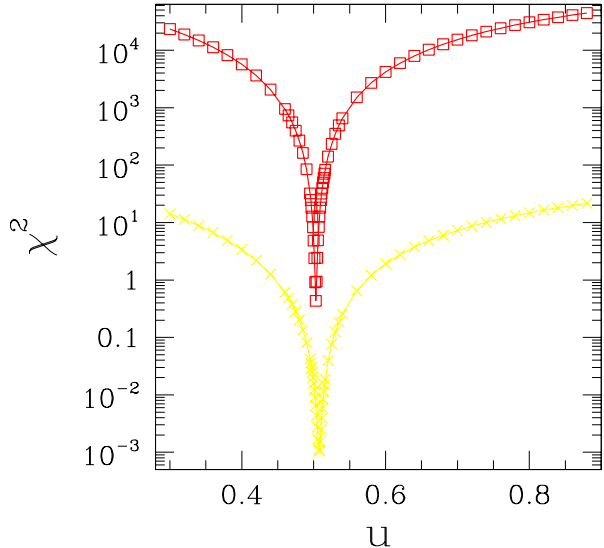}} \par}
\caption{Semilogarithmic plot of $\chi^2$ against $u$ for least-squares
fits of Eq.(\protect{\ref{eq:spfit}}).
Crosses: $\eta_1$; squares: $\eta_2$. Data for $\eta_2$ shifted upwards by 
a factor of 100 on vertical scale, to avoid superposition. For each
fit, only data in the interval $3.05176E-10 \leq x \leq  2.5E-06$ of 
Table~\protect{\ref{table3}} were used (see text).}
\label{fig:spfit}
\end{figure}

Our first test is a single-power fit to scaling corrections: we
assume
\begin{equation}
\eta_i(x)-\eta_i(0)=C_i\,x^u\ ,
\label{eq:spfit}
\end{equation} 
and vary $u$ within a reasonably broad range, checking the behaviour of 
the $\chi^2$ of the corresponding least-squares fit. Our results, using
as input the upper half of Table~\ref{table3} (14 data plus
the $x=0$ line, for each fit) are
displayed in Figure~\ref{fig:spfit}, where very sharp minima can be seen
slightly above $x=0.5$ (to three decimal places, they are located 
respectively at $x=0.508$ for $\eta_1$, $x=0.503$ for $\eta_2$). 
This signals that (i) corrections depending on $\sqrt{x}$ are 
definitely present, thus supporting the assumption made in 
Eq.~(\ref{eq:bcorrx}); and (ii) higher-order terms are not negligible.
Indeed, inclusion of data for larger $x$ causes the $\chi^2$
to increase steeply, while the sharpness of the dips deteriorates, and 
their position shifts towards larger $u$.   

\begin{figure}
{\centering \resizebox*{3.4in}{!}{\includegraphics*{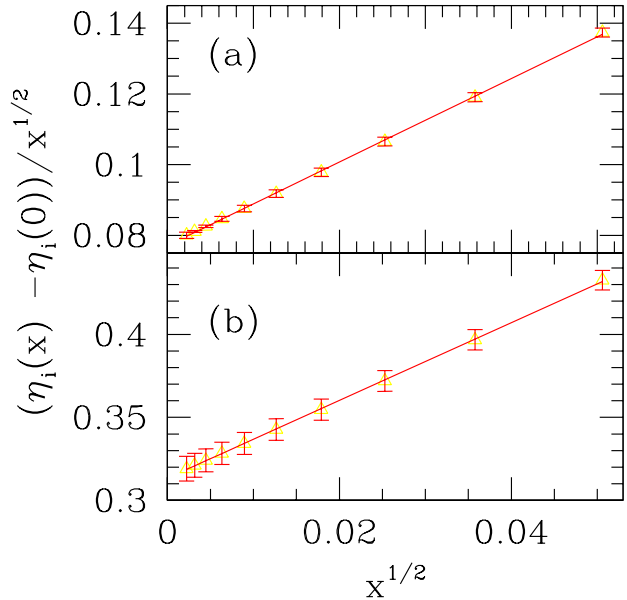}} \par}
\caption{Plots of  $(\eta_i(x)-\eta_i(0))/\sqrt{x}$
against $\sqrt{x}$, for $i=1$ (a) and $i=2$ (b). Straight lines are
linear least-squares fits to the subset of data in plot, corresponding
to  $ 5.E-06 < x < 0.00256$ in
Table~\protect{\ref{table3}} (see text).}
\label{fig:tpfit}
\end{figure}

Having ensured that square-root corrections to scaling are an 
essential element of the picture,
we attempt to include higher-order terms, in the manner of
Eq.~(\ref{eq:bcorrx}). We plot $(\eta_i(x)-\eta_i(0))/\sqrt{x}$
against $\sqrt{x}$, thus one expects:
\begin{equation}
\frac{\eta_i(x)-\eta_i(0)}{\sqrt{x}} = C_i + D_i\,{\sqrt{x}} + {\cal O}(x)
\label{eq:tpfit}
\end{equation}
and attempts  straight-line fits. Results are in 
Figure~\ref{fig:tpfit}.  
The subset of data considered now is complementary to that used in the 
earlier single-power fits, as higher-order terms become more important
for $x$ not very small. We noticed that inclusion of data from the last 
3--4 lines of Table~\ref{table3} caused a quick deterioration of the 
quality of linear fits (the resulting curvature can be seen by naked eye);
this is probalbly the effect of third- and higher-order terms in 
$\sqrt{x}$. An alternative source of errors would be the 
multiplicative logarithmic terms, mentioned in connection with
Eq.~(\ref{eq:bcorrx}) above, and not considered in the present approach. 
Therefore, we decided to keep to the range of $x$ for which
good linear plots were obtainable, while using as many data as possible 
(in order to reduce the spread in the estimates of $C_i$ and $D_i$ of 
Eq.~(\ref{eq:bcorrx})). The best compromise was found by taking data
for  $ 5.E-06 < x < 0.00256$, shown in Figure~\ref{fig:tpfit} together 
with the corresponding least-squares fits.

\begin{table}
\caption{Estimates of amplitudes $C_i$, $D_i$ (see Eqs.~(\ref{eq:bcorrx})
--(\ref{eq:tpfit})), and $\chi^2$ per degree of freedom
($\chi^2/{\rm DOF})$ for respective fits. 
}
\vskip 0.1cm
 \halign to \hsize{\hskip1truecm\hfil#\hfil&\hfil#\hfil&
\hfil#\hfil&\hfil#\hfil\cr
     \ \      &\ $C_i$ &\ $D_i$ &\  $\chi^2/{\rm DOF}$ \cr
Eq.~(\protect{\ref{eq:spfit}}), i=1, u=0.5 &\ $0.079(4)$ &\ $-$  &\ 
$1.5 \times 10^{-3}$ \cr
Eq.~(\protect{\ref{eq:tpfit}}), i=1  &\ $0.077(1)$ &\ $1.18(3)$  &\ 
$2 \times 10^{-2}$ \cr
Eq.~(\protect{\ref{eq:spfit}}), i=2, u=0.5 &\ $0.317(5)$ &\ $-$  &\ 
$3.0 \times 10^{-3}$ \cr
Eq.~(\protect{\ref{eq:tpfit}}), i=2  &\ $0.313(3)$ &\ $2.3(1)$  &\ 
$2 \times 10^{-3}$ \cr}
\label{table4}
\end{table}

Our estimates of the amplitudes $C_i$ and $D_i$ are shown in 
Table~\ref{table4}, where for single-power
fits of Eq.~(\ref{eq:spfit}) we quote only results for $u=0.5$.
Although, as explained above, these do not correspond to the respective 
absolute minima of $\chi^2$, 
they exhibit a very good quality of fit, and it seems more appropriate
to compare them (instead of those obtained at minimal $\chi^2$) to
the estimates of $C_i$ from  Eq.~(\ref{eq:tpfit}), where the power 
$1/2$ is fixed from the start. 

\section{Scaling of third gap}
\label{stg}

Finally, we have investigated the scaling of the third gap at $T=0$. 
According to theory~\cite{dN82,saso98}, at the critical point there are 
only three relevant operators, corresponding (in decreasing order of
relevance) to staggered magnetization, uniform magnetization and
staggered polarization. Although, as recalled above, there is widespread
agreement between theory and numerical work as regards the first two,
the prediction of Ref.~\onlinecite{saso98}, namely that the corresponding
decay-of-correlations exponent is $\eta_{{\bf P}_{\rm stagg}}=3$, appears
not to have been numerically tested so far. ( In Ref.~\onlinecite{saso98},
Monte Carlo simulations were performed for the respective susceptibility,
which according to the scaling law $\gamma/\nu=2-\eta$ is expected to 
approach a constant, with corrections $\propto L^{-\Delta}$, $\Delta=1$, 
if  $\eta=3$; the approach to a constant was indeed verified, while the
best fit was for $\Delta \simeq 0.75$ instead of unity).  

We have calculated descending eigenvalues of the transfer matrix; it would 
seem plausible to associate the third scaled gap to the staggered 
polarization, especially since only three relevant operators are expected
to come up, and the relationship of the other two to the first two gaps is 
well-established. In order to check self-consistency of our results, we 
used both a standard power-method algorithm, coupled 
with Gram-Schmidt orthogonalization, and a Lanczos scheme. While for
small $L$ and shallow levels (corresponding to eigenvalues $\Lambda_i$, $i 
\leq 3$) both methods gave the same estimates, the Lanczos 
results displayed instabilities for deeper levels and $L \geq 8$. At 
present we are not able to explain such discrepancies. Therefore, we
restrict ourselves to the analysis of the third gap. 

Our results, again displayed in the form $\eta_3=
L\, (\Lambda_0-\Lambda_3)/\pi$, are shown in Table~\ref{table5}. 
In order to gain an unbiased perspective both of the limiting bulk 
value of $\eta_3$ and of the scaling corrections, 
we attempted a single-power extrapolation, $\eta_3(L)=\eta_3 +a_{30}/L^u$
with a variable power $u$, and monitored the variations of the $\chi^2$
of the corresponding fits against $u$. The result was qualitatively
very similar to that displayed for the fits of 
Eq.~(\ref{eq:spfit}) in Fig.~\ref{fig:spfit}: a rather sharp minimum,
located at $u=1.8$ in this case, which gave an extrapolated $\eta_3=
2.00(1)$ (see Table~\ref{table5}; the error bar was calculated by 
considering the estimates on either side of $u_{\min}=1.8$, for which the 
$\chi^2$ becomes one order of magnitude larger than at the minimum). 
Fixing $u=1$, inspired by the prediction of Ref.~\onlinecite{saso98}
for the susceptibility, gave $\eta_3 \simeq 2.11$. Two-power fits
{\it \`a la} Eq.~(\ref{eq:corr0}), using either $L^{-2}$ and $L^{-4}$
or $L^{-1}$ and $L^{-2}$ also gave values between $1.99$ and $2.01$.
There seems to be 
no straightforward way to extrapolate the data of Table~\ref{table5}
to include $\eta_3=3$ . At this point we do not know how to reconcile
our results to the predictions of  Ref.~\onlinecite{saso98}.
\begin{table}
\caption{Finite--$L$ and extrapolated estimates of $\eta_3$. 
The latter is the result of an equal-weight fit of data 
for $L=6$, $8 \cdots 14$ respectively to a single-power ($L^{-u}$) 
correction; the corresponding $u=1.8$ was chosen to minimize the
$\chi^2$ (see text). 
}
\vskip 0.1cm
 \halign to \hsize{\hskip2.8truecm\hfil#\hfil&\hfil#\hfil\cr
     L        &\ $\eta_3 $  \cr
     6        &\  1.74149553  \cr
     8        &\  1.84852826  \cr
     10       &\  1.90108612  \cr
     12       &\  1.93052844  \cr
     14       &\  1.94860339  \cr
     Extr.  &\ 2.00(1)  \cr}
\label{table5}
\end{table}

\section{Conclusions}
\label{conc}

In summary, we have undertaken a finite-size approach to investigate
the corrections to scaling in the three-state square-lattice Potts 
antiferromagnet. Owing to the characteristics of the quantities under 
study, we argued that the natural variable to consider is $x \equiv 
L\,e^{-2\beta}$. We showed that the less-relevant finite-size corrections
could be accounted for in a phenomenological scheme, based on the 
zero-temperature picture; for the extrapolated scaled gaps we supplied
convincing evidence that square-root corrections, in the variable $x$,
are present, and provided estimates for the numerical values of
the amplitudes of the first-- and second--order correction terms, for both
the first and second scaled gaps. It would be interesting if predictions
based on theory could be derived, to be compared with the numerical
values of amplitudes obtained in this work.

We have also investigated the behaviour of the third scaled gap of the 
transfer matrix spectrum, and found an extrapolated value for the 
decay-of-correlations exponent $\eta_3=2.00(1)$. This seems incompatible 
with 
earlier predictions, to the effect that the third relevant operator in the
problem would give $\eta_{{\bf P}_{\rm stagg}}=3$,
corresponding to the staggered polarization.

\begin{acknowledgments}

The author thanks Alan Sokal for many interesting discussions and 
suggestions, and for a critical reading of an early version of the 
manuscript; thanks are also due to the Department of Theoretical Physics
at Oxford, where this  work was initiated, for the hospitality, and
to the cooperation agreement between CNPq and
the Royal Society for funding the author's visit.
Research of S.L.A.d.Q. is partially supported by the Brazilian agencies
CNPq (grant No. 30.1692/81.5), FAPERJ (grants
Nos. E26--171.447/97 and E26--151.869/2000) and FUJB-UFRJ.
\end{acknowledgments}


\begin{thebibliography}{99}

\bibitem{Night82} M.\ P.\ Nightingale and M.\ Schick, J.\ Phys.\ A {\bf
15}, L39 (1982).
\bibitem{rw} H.\ H.\ Roomany and H.\ W.\ Wyld, \prd {\bf 21}, 3341
(1980).
\bibitem{Wang90} J.-S.\ Wang, R.\ H.\ Swendsen, and R.\ Koteck\'y,
\prb {\bf 42}, 2465 (1990).
\bibitem{saleur91} H.\ Saleur, Nucl. Phys. B {\bf 360}, 219 (1991).
\bibitem{Ferreira95} S.\ J.\ Ferreira and A.\ D.\ Sokal, \prb {\bf 51},
6727 (1995).
\bibitem{gof4} S. L. A. de Queiroz, T. Paiva, J. S. de S\'a Martins, and 
R. R. 
dos Santos, \pre {\bf 59}, 2772 (1999).
\bibitem{feso99} S.\ J.\ Ferreira and A. D. Sokal, J. Stat. Phys. {\bf 
96}, 461  (1999).  
\bibitem{cjs01} J. L. Cardy, J. L. Jacobsen, and A. D. Sokal, J. Stat. 
Phys. {\bf 105}, 25  (2001).  
\bibitem{cardy} J. L. Cardy, J. Phys. A {\bf 17}, L385 (1984).
\bibitem{dN82} M.\ P.\ M.\ den Nijs, M.\ P.\ Nightingale and M.\ Schick,
\prb {\bf 26}, 2490 (1982) .
\bibitem{saso98} J. Salas and  A. D. Sokal, J. Stat. Phys. {\bf
92}, 729  (1998).
\bibitem{barber}
M. N. Barber, in {\it Phase Transitions and Critical Phenomena},
edited by C. Domb and J.L. Lebowitz (Academic, New York, 1983), Vol. 8.
\bibitem{car86} J. L. Cardy, Nucl. Phys. B {\bf 270}, 186 (1986).
\bibitem{nien82} B. Nienhuis, J. Phys. A {\bf 15}, 199 (1982).
\bibitem{sldq00} S. L. A. de Queiroz, J. Phys. A {\bf 33}, 721 (2000).


%\bibitem{bdn88} H.\ J.\ W.\ Bl\"ote and M.\ P.\ M.\ den Nijs, \prb {\bf
%37}, 1766 (1988).

\end{thebibliography}
\end{document}